\documentclass[prl,twocolumn,superscriptaddress]{revtex4}
\usepackage{epsfig}
\usepackage{times}
\usepackage{amsmath}
\usepackage{amsfonts}
\usepackage{amssymb}
\usepackage[usenames]{color}
\usepackage{graphicx}

\newcommand{\beq}{\begin{equation}}
\newcommand{\eeq}{\end{equation}}
\newcommand{\beqa}{\begin{eqnarray}}
\newcommand{\eeqa}{\end{eqnarray}}

\newcommand{\NTT}{NTT Basic Research Laboratories, NTT Corporation, 3-1 Morinosato-Wakamiya, Atsugi, Kanagawa, 243-0198, Japan.}

\newcommand{\TOKYO}{Department of Physics, Graduate School of Science, The University of Tokyo, Hongo 7-3-1, Bunkyo-ku, Tokyo 113-0033, Japan}
\newcommand{\NII}{National Institute of Informatics, 2-1-2 Hitotsubashi, Chiyoda-ku, Tokyo 101-8430, Japan.}
\begin{document}
\title{Projective measurement of energy \textcolor{black}{on an ensemble of qubits with
unknown frequencies}
}

\author{Yuichiro Matsuzaki}
\email{matsuzaki.yuichiro@lab.ntt.co.jp} \affiliation{\NTT}
\author{Shojun Nakayama} 	\affiliation{\NII}
\author{Akihito Soeda} 	\affiliation{\TOKYO}
\author{Mio Murao} 	\affiliation{\TOKYO}
\author{Shiro Saito} 	\affiliation{\NTT}
\begin{abstract}
 In projective measurements of energy, a target system is projected
 \textcolor{black}{to an} 
 eigenstate of the \textcolor{black}{system} Hamiltonian, and the measurement \textcolor{black}{outcomes} provide
  the information of \textcolor{black}{corresponding eigen-energies}. 
 Recently, it has been shown that such a measurement can be in principle
 \textcolor{black}{realized} without detailed knowledge of the Hamiltonian by using probe qubits.
  However, in the previous approach for the
  energy measurement, the necessary
 \textcolor{black}{size of the dimension for the probe}
  increases as we increase \textcolor{black}{the dimension} of the target system, and also individual addresibility of every qubit is
 required, which may not \textcolor{black}{be possible
 for many experimental settings with large systems.}
 Here, we show that a single probe qubit is sufficient to perform
 such a projective measurement of energy if the target system is
 composed of non-interacting qubits
 whose resonant frequencies are unknown.
 Moreover,
 our scheme requires only global
 manipulations where every qubit is subjected
\textcolor{black}{to} the same control fields.
  These results indicate the feasibility of \textcolor{black}{our} energy
 projection \textcolor{black}{protocols}.
\end{abstract}

\maketitle

 A projective measurement is a central concept in quantum physics \cite{vNbook32springer,koshino2005quantumshimizu,wiseman2009quantum}. This ideally
projects the state
\textcolor{black}{to } the eigenstate of the measured observable.
The quantum measurement process can be described by an interaction between the
target system and a probe system.  A
correlation with the probe system is generated via the coupling, and a
measurement on the probe system induces the projection on the target
system where the measurement outcomes of the probe are associated with the
eigenvalues of the observable. If the measured observable
is energy of the target system, such an operation is called 
projective measurement of energy (PME).

\textcolor{black}{ PME scheme exists}
if the form of the Hamiltonian is given in
advance \cite{aharonov2002measuring}.
Based on the knowledge of the Hamiltonian, we can engineer the
interaction between the target and probe system.
However, to identify the unknown Hamiltonian, it takes at least
$O(d^2)$
time with quantum tomography
where $d$ denotes the Hilbert space of the system
\cite{chuang1997prescription,poyatos1997complete}, and this grows
exponentially \textcolor{black}{with}
the size of the target system.

Nakayama {\it{et al}} proposed a scheme to perform the PME
of unknown Hamiltonians \textcolor{black}{whose dimension and energy scale
are only known.}
\textcolor{black}{The necessary time is independent from}
the dimension
of the Hamiltonian \cite{nakayama2015quantum}.
Quantum phase
estimation (QPE) \textcolor{black}{algorithm is used} 
to estimate eigenvalues of a given
unitary operator $U$ \cite{kitaev2002classical}, and controlled-swap gates
 between the target system and probe system play a central role 
 \textcolor{black}{implementing} the PME. However, there are no existing \textcolor{black}{schemes} to construct
 the controlled-swap gate \textcolor{black}{in actual experiments} when the
 Hamiltonian is unknown. Moreover, 
 their protocol requires a probe system whose size is comparable with
 that of the target system, while individual
 controllability for every qubit is requisite.
 \textcolor{black}{Due to these restrictions, it is not clear whether
 their protocol could be demonstrated in actual experiments.}

In this letter, we introduce a scheme to implement the PME on
\textcolor{black}{an ensemble of qubits with
unknown frequencies}
where only
\textcolor{black}{global control with a single probe qubit is} required, while keeping the advantage of the
\textcolor{black}{reduced} time cost.
We \textcolor{black}{consider} that 
a single probe qubit is
collectively coupled with \textcolor{black}{the target qubits where interaction
between the target qubits is negligible}.
Without detailed knowledge of the \textcolor{black}{target qubits},
\textcolor{black}{one can} perform the PME via the coupling
with the probe qubit, \textcolor{black}{while the necessary time cost is independent from}
the dimension of the Hamiltonian. Moreover, our protocol just requires global controls where
all qubits are subjected in the same external fields.
These advantages show our PME \textcolor{black}{is much more suitable
for experimental realizations than the previous schemes.}

Let us review quantum phase estimation (QPE) \cite{kitaev2002classical}.
Here, we consider the case that QPE is performed on
a given unitary operation under the assumption that
\textcolor{black}{implementation of} a controlled unitary
gate and Fourier basis measurements are available.
\textcolor{black}{By QPE,} we can estimate eigenvalues of
a target unitary operator
$U_t=e^{-iH_{\rm{QPE}}t}=\sum_{n=0}^{2^N-1}e^{-iE_nt}|E_n\rangle \langle
E_n|$
where
$|E_n\rangle $  \textcolor{black}{and} $E_n$ denote eigenvectors
\textcolor{black}{and eigenvalues} of 
\textcolor{black}{the target Hamiltonian}
$H_{\rm{QPE}}$, \textcolor{black}{respectively}. We assume $0\leq  E_nt
<2\pi $ $(n=1,2,\cdots ,L)$ to remove an ambiguity due
to a phase periodicity.
A control-unitary operation $\mathcal{C}_{U_t}=|0\rangle _{\rm{P}}\langle
0|\otimes \openone_{\rm{T}}+|1\rangle _{\rm{P}}\langle 1|\otimes
U_t $
between the probe qubit and the target qubits is required 
for the implementation of QPE. This operation can induce 
a phase kick back $\mathcal{C}_{U_t} |+\rangle _{\rm{P}}
|E_n\rangle _{\rm{T}}=\frac{1}{\sqrt{2}}(|0\rangle _{\rm{P}}+e^{-iE_nt}|1\rangle
_{\rm{P}} ) |E_n\rangle _{\rm{T}}$, which is \textcolor{black}{essential
for QPE.}
The QPE exploits $L$ probe qubits,
and we apply $L$ control-unitary operations to obtain
$\mathcal{C}^{(1)}_{U_{t}}\mathcal{C}^{(2)}_{U_{2t}}\cdots
\mathcal{C}^{(L)}_{U_{2^{L-1}t}}(\bigotimes _{j=1}^L|+\rangle _{\rm{P}_j})|\phi
\rangle _{\rm{T}} $ where $\mathcal{C}^{(j)}_{U_{2^{j-1}t}}$ is performed
between \textcolor{black}{the} $j$ th probe qubit and the target qubits. By measuring the probe
qubits \textcolor{black}{in the computational basis}
after
the
quantum Fourier transform, 
the probe qubits \textcolor{black}{are measured}
in the Fourier basis
\begin{eqnarray}
 |m\rangle _{\rm{P}}
  =\frac{1}{2^{\frac{L}{2}}}\sum_{k_1=0}^{1}\sum_{k_2=0}^{1}\cdots
  \sum_{k_L=0}^{1}
  \bigotimes _{j=1}^L
  e^{2\pi i \cdot 2^{L-j}k_j m/2^L}|k_j\rangle _{\rm{P}_j}\ 
\end{eqnarray}
where $m=0,1,\cdots , 2^L-1$.
With a limit of a large $L$, the measurement
\textcolor{black}{outcomes} $m$ correspond to the values of $E_nt$ such as
$\frac{E_nt}{2\pi }=\frac{m}{2^L}$, and
these phases \textcolor{black}{can be estimated}
by QPE.
\textcolor{black}{Thus,} the QPE is equivalent to the PME
\textcolor{black}{of $H_{\rm{QPE}}$}. 
We can replace the $L$ probe qubits with a
single probe qubit for \textcolor{black}{performing} the Fourier basis measurement, \textcolor{black}{if
the controlled unitary gate is given}
\cite{griffiths1996semiclassical,parker2000efficient}.
\textcolor{black}{In this case, }
we need to reset, rotate, and measure the probe qubit $L$ times \textcolor{black}{using} a technique of
measurement feedback where the angle of the rotations depends on
previous measurement \textcolor{black}{outcomes} \cite{griffiths1996semiclassical,parker2000efficient}.

The most difficult part to realize the PME is to construct the
controlled unitary gate for an unknown Hamiltonian.
Here, we propose a way that approximately
implements such a controlled unitary gate
with a limited knowledge of the Hamiltonian by using a single probe qubit.

We consider a system where the probe qubit is collectively coupled with the target qubits
and the microwave fields are globally coupled with the qubits.
We assume that
an interaction \textcolor{black}{among} the target qubits is negligible.
The \textcolor{black}{joint} Hamiltonian of the probe and target systems
\textcolor{black}{is given by}
\begin{eqnarray}
 &&H=\frac{\omega _{\rm{P}}}{2}\hat{\sigma }^{\rm{(P)}}_z+\lambda
  _{\rm{P}}\cos (\omega t) \hat{\sigma }_x^{\rm{(P)}}
  \nonumber \\
  &+&\sum_{j=1}^{N}\Big{(}\frac{g}{2}\hat{\sigma
  }^{\rm{(P)}}_z
  \hat{\sigma }^{\rm{(T)}}_{z,j}+ \frac{\omega _{j}}{2}\hat{\sigma }^{\rm{(T)}}_{z,j} +\lambda_{\rm{T}} \cos (\omega 't)
  \hat{\sigma }^{\rm{(T)}}_{x,j}\Big{)} 
\end{eqnarray}
where $\omega _j$ denotes the
frequency of the $j$-th target qubit, $\omega _{\rm{P}}$ denotes the frequency of the probe
qubit, $g$ denotes a coupling strength, $\lambda _{\rm{P}}$ ($\lambda
_{\rm{T}}$) denotes the Rabi frequency for the probe (target) system,
$\omega $ ($\omega '$) denotes the frequency of the microwave for the probe (target) system.
\textcolor{black}{We aim} to realize PME of the target Hamiltonian
$H_{\rm {T}}=\sum_{j=1}^{N}\frac{\omega _{j}}{2}\hat{\sigma
}^{\rm{(T)}}_{z,j} $.
We assume 
the average frequency
$\omega _{\rm{av}}$
and the
variance $\delta \omega $ of the target qubits are given, but the individual
frequency $\omega _j$ is unknown.
By detuning
the probe qubit frequency
from the average frequency of the target qubits,
  we can
  control the probe qubit without affecting the target qubits. 
  In a rotating frame,
  we rewrite
  the Hamiltonian
  with
  rotating wave approximation
  \textcolor{black}{as}
  \begin{eqnarray}
 H&\simeq &\frac{\lambda _{\rm{P}}}{2}\hat{\sigma
  }^{(\rm{P})}_x+\sum_{j=1}^{N}\frac{g}{2}
  (\hat{\openone}+\hat{\sigma}^{\rm{(P)}}_z)
  \hat{\sigma }^{\rm{(T)}}_{z,j}\nonumber \\
  &+&\sum _{j=1}^N\Big{(} \frac{\delta \omega _{j}}{2}\hat{\sigma }^{\rm{(T)}}_{z,j} +\frac{\lambda_{\rm{T}} }{2}
  \hat{\sigma }^{\rm{(T)}}_{x,j}\Big{)}
  \label{hamiltonian}
\end{eqnarray}
where $\omega =\omega _{\rm{P}}$, $\omega '=\omega _{\rm {av}}-g$, and
$\delta \omega _j=\omega _j-\omega _{\rm{av}}$. 
We assume
a tunability to turn on/off $\lambda _{\rm{P}}$, $\lambda _{\rm{T}}$,
and $g$.
\textcolor{black}{From 
Eq. (\ref{hamiltonian}),
we define
$H_{\pm }$  by substituting $\lambda _{\rm{P}}=0$ and $\lambda
_{\rm{T}}=\pm \lambda $  while we define $H_{\rm{0}}$
by substituting} $\lambda _{\rm{P}}=\lambda _{\rm{T}}=g=0$.

We show that
it is possible to
construct an approximate controlled-not
(CNOT) gate between the probe and unknown target qubits.
When the probe qubit state is
  $|0\rangle _{\rm {P}}$ ($|1\rangle _{\rm {P}}$) for $H_{\pm }$,
the Hamiltonian of the $j$th target qubit is
represented as $H^{(\pm )}_{j,|0\rangle _{\rm {P}}}=\frac{\delta \omega
_{j}}{2}\hat{\sigma }^{\rm{(T)}}_{z,j} \pm \frac{\lambda }{2}
  \hat{\sigma }^{\rm{(T)}}_{x,j}$ ($H^{(\pm )}_{j,|1\rangle _{\rm
  {P}}}=\frac{\delta \omega _{j}+2g}{2}\hat{\sigma }^{\rm{(T)}}_{z,j}
  \pm \frac{\lambda }{2}
  \hat{\sigma }^{\rm{(T)}}_{x,j} $). 
   We obtain
   \textcolor{black}{$|{}_j\langle 1|e^{-iH^{(\pm )}_{j,|0\rangle _{\rm {P}}}\frac{\pi }{\lambda}}
   |0\rangle _j|^2\simeq 1-\frac{3}{4}|\frac{\delta \omega
  _j}{g}|^2+O(|\frac{\delta \omega
  _j}{g}|^3)$ and $|{}_j\langle 0|e^{-iH^{(\pm )}_{j,|1\rangle _{\rm {P}}}\frac{\pi }{\lambda}}
  |0\rangle _j|^2= |{}_j\langle 1|
  e^{-iH^{(\pm )}_{j,|1\rangle _{\rm {P}}}\frac{\pi }{\lambda}}
  |1\rangle _j|^2\simeq 1-\frac{9\pi ^2}{256}|\frac{\delta \omega
  _j}{g}|^2+O(|\frac{\delta \omega
  _j}{g}|^3)$   } 
   where $\lambda
  =\frac{2}{\sqrt{3}}g$. The target qubit is approximately
  flipped (unchanged) if the probe qubit is $|0\rangle _{\rm {p}}$
  ($|1\rangle _{\rm {p}}$).
  \textcolor{black}{Thus,} $e^{-iH_{\pm }\frac{\pi }{\lambda }}$
  \textcolor{black}{corresponds to} an \textcolor{black}{approximated} CNOT
  gate  up to local operations.

  If such an approximated CNOT gate is given, it is straightforward
  to implement a control unitary gate \textcolor{black}{acting on} the probe qubit and
  unknown target
  qubits.
   We assume that
   the target system is in a
  thermal equilibrium state $\rho _{\rm{T}} = \frac{1}{Z}
  e^{-\frac{H_T}{k_BT_E}}$ where $k_B$ denotes \textcolor{black}{the} Boltzmann constant, 
  $T_E$ denotes an environmental temperature, and $Z$ denotes a partition
  function $Z={\rm{Tr}}[e^{-\frac{H_{\rm{T}}}{k_BT_E}}]$. 
  The target
  system \textcolor{black}{can be interpreted as} a classical mixture of $\bigotimes _{j=1}^N|s_j\rangle_{{\rm {T}}_j}$
  where $s_j=0,1$ $(j=1,2,\cdots ,N)$, and we consider the case that
  the initial state is one of these states. 
  Firstly, prepare the state $|+\rangle _{\rm {P}} \bigotimes
  _{j=1}^N|s_j\rangle _{{\rm {T}}_j}$. Secondly, perform
  \textcolor{black}{the approximated} CNOT gate
  \textcolor{black}{on the state},
  and obtain $\frac{1}{\sqrt{2}}|0\rangle _{\rm {P}} \bigotimes
  _{j=1}^N|s_j\rangle _{{\rm {T}}_j} + \frac{1}{\sqrt{2}}|1\rangle _{\rm {P}} \bigotimes
  _{j=1}^N|\overline{s_j}\rangle _{{\rm {T}}_j}$ where $\hat{\sigma
  }^{({\rm {T}})}_{x,j}|s_j\rangle _{{\rm {T}}_j}
  =|\overline{s_j}\rangle _{{\rm {T}}_j} $.
  Thirdly, let this state evolve with a Hamiltonian $H_{\rm {0}}$,
  \textcolor{black}{to} obtain $\frac{1}{\sqrt{2}}|0\rangle _{\rm {P}} \bigotimes
  _{j=1}^N|s_j\rangle _{{\rm {T}}_j} + \frac{1}{\sqrt{2}}e^{i\sum_{j=1}^{N}\delta \omega
  _jt }
  |1\rangle _{\rm {P}} \bigotimes
  _{j=1}^N|\overline{s_j}\rangle _{{\rm {T}}_j}$.
  \textcolor{black}{Although the probe qubit may suffer
  from decoherence
  during this time evolution,
  we could use a quantum
  memory with a longer coherence time to store the probe state only during
  the free evolution.}
  Finally, by performing
  the second \textcolor{black}{approximated} CNOT gate, we obtain
  $\frac{1}{\sqrt{2}}(|0\rangle _{\rm {P}}+ e^{i\sum_{j=1}^{N}\delta \omega
  _jt }|1\rangle _{\rm {P}}) \bigotimes
  _{j=1}^N|s_j\rangle _{{\rm {T}}_j}$.
  This
  \textcolor{black}{implements a}
  controlled unitary gate
  \textcolor{black}{on} the probe qubit and unknown
  target qubits.

  \begin{figure}[h] 
\begin{center}
\includegraphics[width=1.03\linewidth]{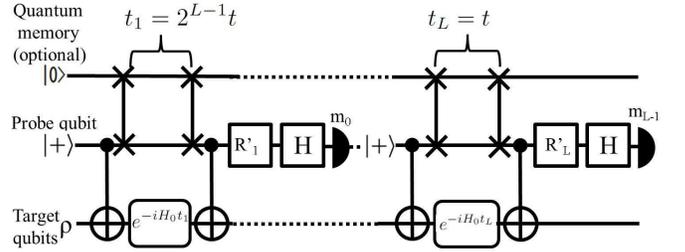} 
\caption{
 \textcolor{black}{A quantum circuit representing our protocol}
 for PME. 
 A combination of the CNOT gates and
 free time evolution \textcolor{black}{for a time $t$} 
 provides the
 probe qubit with a phase information of the target systems.
 To avoid decoherence on the probe qubit, we perform a SWAP gate
 between the probe qubit and memory qubit (with a long coherence time)
  before and after the free evolution of the target system.
 To perform a Fourier basis measurement, we reset, rotate, and measure
 the probe qubit $L$ times where
 $R_j$ denotes the \textcolor{black}{unitary} rotation whose angle is defined by
 \textcolor{black}{the} previous measurement \textcolor{black}{outcomes}: $R_j'=|1\rangle \langle 1|-\phi
 _j'|0\rangle \langle 0|$ with $\phi '_j={\rm{Exp}}[-2\pi i
 \sum_{k=2}^{j}m_{j-k}/2^k]$.
 The  measurement \textcolor{black}{outcomes} reveal the energy
 \textcolor{black}{eigenvalues} of the target qubits,
 and induce the energy projection.
 }
 \label{schematic}
\end{center}
\end{figure}

  By combining the controlled unitary gate and Fourier basis measurements,
  the PME \textcolor{black}{is implementable}.
  \textcolor{black}{A quantum circuir representing our protocol is shown
  in the}
  Fig. \ref{schematic}. 
  \textcolor{black}{To} perform the Fourier basis measurement, we recycle the single
  probe qubit \textcolor{black}{by using the method}
  proposed in \cite{griffiths1996semiclassical,parker2000efficient}. We measure the
  probe qubit $L$ times, and obtain measurement \textcolor{black}{outcomes} $\{m_j\}^{j=L}_{j=1}$.
  From these, \textcolor{black}{eigenvalues of the target can be estimated
  as}
  $\sum_{j=1}^{N}\delta \omega _j\cdot (s_j-\frac{1}{2})\simeq 2^{-L}m\frac{\pi }{t}$
  where $m=\sum_{j=1}^{L}2^{L-j}m_{j-1}$, and this operation projects
  the target state to one of the energy eigenstates.
  The Kraus operator of our PME \textcolor{black}{protocol} on the target qubit is calculated as
  \begin{eqnarray}
   \hat{V}_m=\sum_{s_1,\cdots, s_L=0,1}e^{-2\pi i
    \frac{m}{2^L}\sum_{j=1}^{L}2^{L-j}s_j}\bigotimes _{j=1}^N
    \prod_{k=1}^L
    U_{s_k,2^{L-k}t}^{(j)}\ \ 
  \end{eqnarray}
  where
  $U_{s,t}^{(j)}=e^{-iH^{(-)}_{j,|s\rangle
  _{\rm{P}}}\frac{\pi }{\lambda }}e^{-iH_{\rm{0}}t}
  e^{-iH^{(+)}_{j,|s\rangle _{\rm{P}}}\frac{\pi }{\lambda }}$.
The probability to obtain the
  measurement \textcolor{black}{outcome} $m$ is $P_m = (\bigotimes
  _{j=1}^N{}_{T_j}\langle s_j| )\hat{V}_m^{\dagger }\hat{V}_m (\bigotimes
  _{j=1}^N|s_j\rangle _{T_j})$ and the post-measurement state of the
  target qubits is
  described as $|\psi _m\rangle _{T}=\frac{1}{\sqrt{P_m}}\hat{V}_m (\bigotimes
  _{j=1}^N|s_j\rangle _{T_j}) $. For a quantum non-demolition
  measurement of energy \cite{braginsky1996quantum},
 an average fidelity
  $F=\sum_{m}P_mF_m $ should be unity where $F_m=|{}_{T}  \langle \psi _m|(\bigotimes
  _{j=1}^N|s_j\rangle _{T_j}) |^2$. We define $\epsilon=
  1-F$ as a projection error of the PME \textcolor{black}{protocol}.

We discuss possible physical systems to realize our protocol. Nitrogen
vacancy (NV) centers in diamond are one of the candidates.
We can
control
the NV center by applying microwave pulse, and also readout the
spin state of the NV center by an optical detection \cite{doherty2013nitrogen}.
The NV center
is coupled with nuclear spins via a hyperfine couplings
\cite{NMRHWYJGJW01a,shimo2015control}.
\textcolor{black}{We could
implement our PME on the nuclear spins by using the NV center as a probe
qubit.}
Superconducting circuits are also promising candidates. Recently,
a coherent coupling between a superconducting qubit ensemble and a
microwave resonator has been demonstrated \cite{macha2014implementation,kakuyanagi2016observation},
\textcolor{black}{
and
our PME is implementable on the superconducting qubits via the microwave cavity}
if we
control the microwave cavity as an effective two-level system by using a Kerr effect
\cite{heeres2015cavity}.

\textcolor{black}{Among many candidates, we especially focus on a 
superconducting flux qubit coupled with an electron spin ensemble
\cite{ClarkeWilhelm01a,marcos2010coupling,twamley2010superconducting}. }
Here,
we could implement \textcolor{black}{our PME protocol} on the electron spins by using the flux qubit as a probe.
High fidelity control and readout of the superconducting flux qubit have
been demonstrated \cite{ClarkeWilhelm01a}.
Recently, a coherent
coupling between the flux qubit and the electron spin ensemble was
observed, and the coupling strength between a single electron spin and a
flux qubit is estimated as $g/2\pi =10$ kHz \cite{zhu2011coherent}. Moreover, there is a theoretical proposal to increase \textcolor{black}{coupling} up
to $g/2\pi =100$ kHz \cite{twamley2010superconducting}.
Although the coherence time of the flux qubit is \textcolor{black}{still} around 80 micro
seconds \cite{yan2015flux} which may not be long enough to realize the
PME \textcolor{black}{protocol}, a quantum
memory \textcolor{black}{for the superconducting qubit} with much longer
coherence time
such as microwave cavitys and solid state spin systems can be used
\cite{marcos2010coupling,twamley2010superconducting,tyryshkin2012electron,reagor2015quantum}.
Especially,
if we can use the nuclear spins
for the quantum memory of the flux qubit, the coherence time can be an
order of an hour \cite{saeedi2013room}. 
In this letter,  we especially consider these
 systems. 

 \begin{figure}[h] 
\begin{center}
\includegraphics[width=0.94\linewidth]{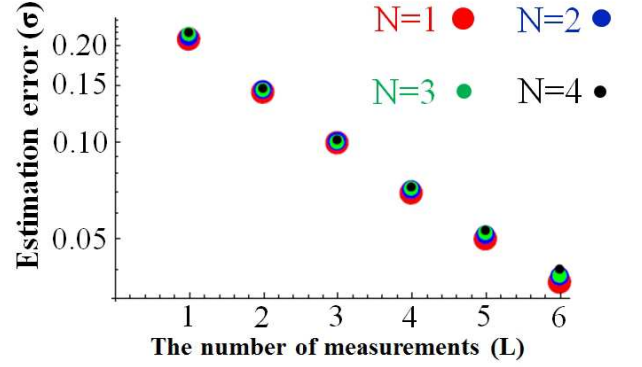} 
\caption{The
 variance ($\sigma $) of
 estimation against the number of measurements ($L$).
  The parameters are $g/2\pi =100$ kHz, $t\sqrt{N}=0.16$ ms, and $\sigma
 _{\rm{G}}/2\pi =1$kHz where $N$ denotes the number of the target
 qubits. \textcolor{black}{In our scheme, $\sigma $ decreases
exponentially with $L$.}
 }
 \label{quadratic}
\end{center}
 \end{figure}
We investigate the performance of the PME \textcolor{black}{protocol} where a single target qubit is
coupled with a probe
qubit. We assume that the initial state of
the target qubit is $|0\rangle _{\rm {T}_1}$ where the detuning of the
target qubit $\delta \omega _1$ has a Gaussian distribution with a zero
average and a variance of $\sigma _{\rm {G}}$. The performance of
our PME \textcolor{black}{protocol} depends on the value of $\delta \omega _1$. \textcolor{black}{To}
evaluate the average performance of our protocol, we
randomly pick up $N_{\rm {r}}$ values of the detuning $\{\delta \omega ^{(l)}_1
\}^{l=N_{\rm {r}}}_{l=1}$ from the Gaussian distribution,
and we will
take an average.

We numerically calculate
the variance of the
\textcolor{black}{estimated energy eigenvalues in our protocol.
The variance is defined as }
\begin{eqnarray}
 \sigma
  &=&\frac{1}{N_{\rm{r}}}\sum_{l=1}^{N_r}\sqrt{\sum_{m=0}^{2^{L-1}}P^{(l)}_m(f_m
  -\frac{\delta \omega ^{(l)}_1t}{2\pi })^2} \label{variance}
\end{eqnarray}
where $P^{(l)}_m$ denotes a probability to obtain a measurement \textcolor{black}{outcome} of
$m$ for a given $\delta \omega ^{(l)}_1$.
The function $f_m=(2^{-L}m-1 )\cdot H_{2^{-L}m- 0.5}+2^{-L}m(1- H_{2^{-L}m-
  0.5}) $ plays a role to remove the ambiguity due
to the phase periodicity where $H_x$ denotes a Heaviside step function.
We plot the variance by the simulations
in Fig. \ref{quadratic}.
The variance decreases
exponentially
\textcolor{black}{with} the number of measurements
$\sigma \propto 2^{-\frac{L}{2}}$,
which is consistent with the scaling of the typical QPE protocol
\cite{higgins2007entanglement,berry2009perform}.

    \begin{figure}[t] 
\begin{center}
\includegraphics[width=1.0\linewidth]{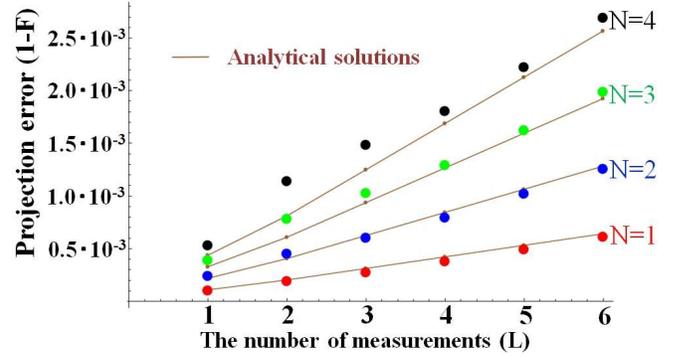} 
\caption{
 The projection errors $(1-F)$
 against the number of measurements $(L)$ \textcolor{black}{where $F$
 denotes a fidelity of the post measurement states}.
 \textcolor{black}{The projection
 error increases linearly with the number of the target qubits $N$.}
 We use the same parameters as \textcolor{black}{in} the Fig. \ref{quadratic}.
 }
 \label{manyfidelity}
\end{center}
\end{figure}
We calculate
the average fidelity between the initial state and the post-measurement
state of the single
target qubit.
\textcolor{black}{We define the fidelity as $F=\frac{1}{N_{\rm{r}}}\sum_{l=1}^{N_r}\sum_{m=0}^{2^{L-1}}P^{(l)}_mF^{(l)}_m$}
where $P^{(l)}_m$ ($F_m^{(l)}$  )
  denote the probability (fidelity) for a measurement \textcolor{black}{outcome} of $m$
  for a given $\delta \omega ^{(l)}_1$.
  If the fidelity is close to the unity,
  we obtain
  an analytical solution of
  the projection error
  \begin{eqnarray}
   \epsilon \simeq \sum_{n=0}^{L-1}\frac{3(64+ 3\pi ^2 +
    e^{-\frac{1}{2}\sigma ^2_{\rm{G}} (2^nt)^2}(64-3\pi ^2)(1- \sigma
    ^2_{\rm{G}} (2^nt)^2 ))}{256(\frac{g}{\sigma _{\rm
    {G}}})^2} \nonumber
  \end{eqnarray}
  We plot the projection error  of both numerical simulations and analytical results
in Fig. \ref{manyfidelity}. 
\textcolor{black}{This} analytical solutions agree
with the numerical simulations. 

We calculate the estimate variance of \textcolor{black}{our} PME for multiple target
qubits. Without loss of generality, we can assume the initial state of
the target qubits is $\bigotimes _{j=1}^N|0\rangle _{\rm {T}_j}$.
We define the variance for multiple target qubits by Eq. \ref{variance}
where we replace $\delta \omega _1^{(l)}$ with $\delta \omega
_l=\sum_{j=1}^{N}\delta \omega ^{(l)}_j$. Here,  $\delta \omega ^{(l)}_j$
denotes a detuning at the $j$ th target qubit
where we
randomly pick up $N_{\rm {r}}$ values of the detuning $\{\delta \omega ^{(l)}_j
\}^{l=N_{\rm {r}}}_{l=1}$ from a Gaussian distribution with zero
 average and variance of $\sigma _{\rm {G}}$.
 Since  we have
 $\delta \omega t= \frac{1}{N_{\rm{r}}}\sum_{l=1}^{N_{\rm{r}}}\delta
 \omega _l\simeq \sqrt{N} \sigma _{\rm {G}} t$ from the central limit theorem,
  we obtain $\delta \omega
 t=\Theta (N^0)$ by choosing $t=\Theta (N^{-\frac{1}{2}})$. So we can
 remove the $N$ dependency of the
 variance.
 We confirm this from numerical simulations, and plot the results
 in Fig. \ref{quadratic}. Similar to the single target qubit case, 
 we can exponentially suppress the
 estimation variance  for
 multiple target qubits as we increase the number of the measurements.

 We also calculate the projection error for multiple target qubits. For a
 single target qubit, we obtained an analytical solution of the projection
 error $\epsilon $. For a small projection
 error, the total projection error for $N$ target qubits will
 be approximated as
 $ N\epsilon $. We plot the analytical
 solution and numerical results in the Fig. \ref{manyfidelity}, and
 there is a good agreement between the analytical and numerical results.
 Since
  the projection error is proportional to the number of the target
  qubits, our PME works efficiently for a relatively small number
  of target qubits.
    For example, from the simulation,
   the projection error is estimated around
    $0.27\%$ for $N= 4$ and $L= 6$ with the current parameters.
   To decrease the projection
 error for a larger number of the target qubits, we should increase the
 coupling strength between the probe and target qubits, which
 would be possible by changing the design of the qubits \cite{twamley2010superconducting,yoshihara2016superconducting}.

     \begin{figure}[h] 
\begin{center}
\includegraphics[width=1.0\linewidth]{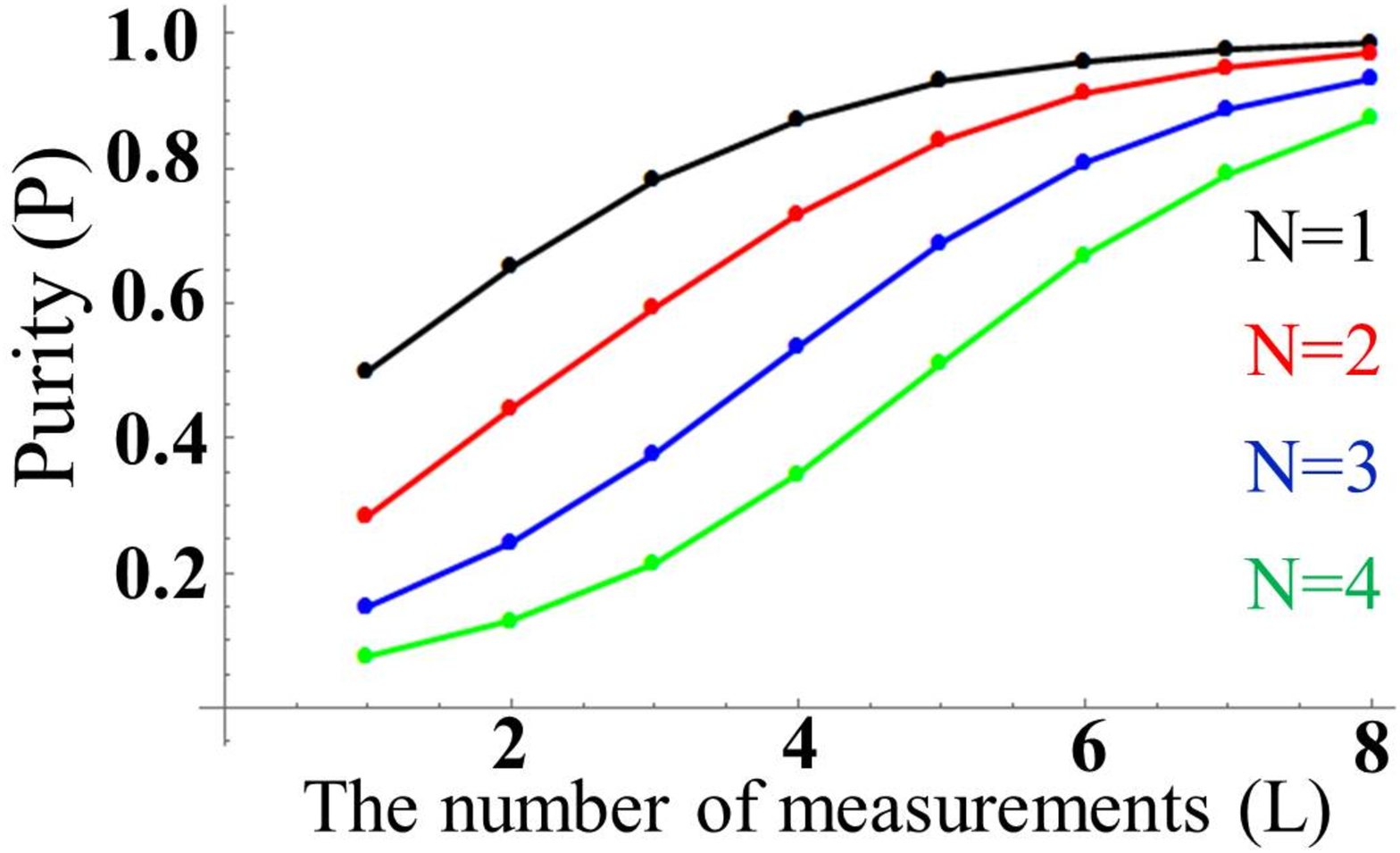} 
\caption{
 The purity of
 the state against the number of measurements.
 We use the same parameters as in the Fig. \ref{quadratic}.
 The purity approaches to the unity as we increase the number of the measurements.
 \textcolor{black}{Each} continuous
 line is a guide for the eyes.
 }
 \label{manypurity}
\end{center}
\end{figure}

 Finally, we calculate the purity
 of the post-measurement state when the
 initial state is a completely mixed state.
 For a given set of the
 detuning $\delta \omega _j^{(l)}$ $(j=1,2,\cdots , N)$, 
 the
 post-measurement state \textcolor{black}{is calculated} as $\rho ^{(l)}_m =\frac{\hat{V}^{(l)}_m
 \frac{1}{2^N}\hat{\openone}(\hat{V}^{(l)}_m)^{\dagger }}{{\rm{Tr}}[\hat{V}_m
 \frac{1}{2^N}\hat{\openone}(\hat{V}^{(l)}_m)^{\dagger }]}$.
 The average purity is calculated as
 $P=\frac{1}{N_{\rm {r}}}\sum_{l=1}^{N_{\rm{r}}}(\rho ^{(l)}_m )^2$.
 We plot this results in Fig. \ref{manypurity}.
 As we increase the number of the measurements, the purity approaches to
 the unity.

In conclusion, we propose a
protocol to implement the projective measurement of energy on
\textcolor{black}{an ensemble of qubits with
unknown frequencies.}
We use a quantum phase estimation algorithm to determine the unknown
energy of the target system.
Unlike previous protocols, we only require a single probe qubit and global
operations for the implementation, which makes it more feasible to
realize.
Our scheme has many potential applications such as characterization of
unknown quantum systems, quantum metrology, and initialization.
This work was supported by JSPS
KAKENHI Grant 15K17732, and was also partially supported by
MEXT KAKENHI Grant Number 15H05870, \textcolor{black}{16H01050,
and 24106009.}

   

\begin{thebibliography}{28}
\expandafter\ifx\csname natexlab\endcsname\relax\def\natexlab#1{#1}\fi
\expandafter\ifx\csname bibnamefont\endcsname\relax
  \def\bibnamefont#1{#1}\fi
\expandafter\ifx\csname bibfnamefont\endcsname\relax
  \def\bibfnamefont#1{#1}\fi
\expandafter\ifx\csname citenamefont\endcsname\relax
  \def\citenamefont#1{#1}\fi
\expandafter\ifx\csname url\endcsname\relax
  \def\url#1{\texttt{#1}}\fi
\expandafter\ifx\csname urlprefix\endcsname\relax\def\urlprefix{URL }\fi
\providecommand{\bibinfo}[2]{#2}
\providecommand{\eprint}[2][]{\url{#2}}

\bibitem[{\citenamefont{von Neumann}(1932)}]{vNbook32springer}
\bibinfo{author}{\bibfnamefont{J.}~\bibnamefont{von Neumann}},
  \emph{\bibinfo{title}{Mathematische Grundlagen der Quantenmechanik}}
  (\bibinfo{publisher}{Springer}, \bibinfo{year}{1932}).

\bibitem[{\citenamefont{Koshino and Shimizu}(2005)}]{koshino2005quantumshimizu}
\bibinfo{author}{\bibfnamefont{K.}~\bibnamefont{Koshino}} \bibnamefont{and}
  \bibinfo{author}{\bibfnamefont{A.}~\bibnamefont{Shimizu}},
  \bibinfo{journal}{Phys. Rep.} \textbf{\bibinfo{volume}{412}},
  \bibinfo{pages}{191} (\bibinfo{year}{2005}).

\bibitem[{\citenamefont{Wiseman and Milburn}(2009)}]{wiseman2009quantum}
\bibinfo{author}{\bibfnamefont{H.~M.} \bibnamefont{Wiseman}} \bibnamefont{and}
  \bibinfo{author}{\bibfnamefont{G.~J.} \bibnamefont{Milburn}},
  \emph{\bibinfo{title}{Quantum measurement and control}}
  (\bibinfo{publisher}{Cambridge University Press}, \bibinfo{year}{2009}).

\bibitem[{\citenamefont{Aharonov et~al.}(2002)\citenamefont{Aharonov, Massar,
  and Popescu}}]{aharonov2002measuring}
\bibinfo{author}{\bibfnamefont{Y.}~\bibnamefont{Aharonov}},
  \bibinfo{author}{\bibfnamefont{S.}~\bibnamefont{Massar}}, \bibnamefont{and}
  \bibinfo{author}{\bibfnamefont{S.}~\bibnamefont{Popescu}},
  \bibinfo{journal}{Phys. Rev. A} \textbf{\bibinfo{volume}{66}},
  \bibinfo{pages}{052107} (\bibinfo{year}{2002}).

\bibitem[{\citenamefont{Chuang and Nielsen}(1997)}]{chuang1997prescription}
\bibinfo{author}{\bibfnamefont{I.~L.} \bibnamefont{Chuang}} \bibnamefont{and}
  \bibinfo{author}{\bibfnamefont{M.~A.} \bibnamefont{Nielsen}},
  \bibinfo{journal}{Journal of Modern Optics} \textbf{\bibinfo{volume}{44}},
  \bibinfo{pages}{2455} (\bibinfo{year}{1997}).

\bibitem[{\citenamefont{Poyatos et~al.}(1997)\citenamefont{Poyatos, Cirac, and
  Zoller}}]{poyatos1997complete}
\bibinfo{author}{\bibfnamefont{J.}~\bibnamefont{Poyatos}},
  \bibinfo{author}{\bibfnamefont{J.~I.} \bibnamefont{Cirac}}, \bibnamefont{and}
  \bibinfo{author}{\bibfnamefont{P.}~\bibnamefont{Zoller}},
  \bibinfo{journal}{Phys. Rev. Lett.} \textbf{\bibinfo{volume}{78}},
  \bibinfo{pages}{390} (\bibinfo{year}{1997}).

\bibitem[{\citenamefont{Nakayama et~al.}(2015)\citenamefont{Nakayama, Soeda,
  and Murao}}]{nakayama2015quantum}
\bibinfo{author}{\bibfnamefont{S.}~\bibnamefont{Nakayama}},
  \bibinfo{author}{\bibfnamefont{A.}~\bibnamefont{Soeda}}, \bibnamefont{and}
  \bibinfo{author}{\bibfnamefont{M.}~\bibnamefont{Murao}},
  \bibinfo{journal}{Phys. Rev. Lett.} \textbf{\bibinfo{volume}{114}},
  \bibinfo{pages}{190501} (\bibinfo{year}{2015}).

\bibitem[{\citenamefont{Kitaev et~al.}(2002)\citenamefont{Kitaev, Shen, and
  Vyalyi}}]{kitaev2002classical}
\bibinfo{author}{\bibfnamefont{A.~Y.} \bibnamefont{Kitaev}},
  \bibinfo{author}{\bibfnamefont{A.}~\bibnamefont{Shen}}, \bibnamefont{and}
  \bibinfo{author}{\bibfnamefont{M.~N.} \bibnamefont{Vyalyi}},
  \emph{\bibinfo{title}{Classical and quantum computation}},
  vol.~\bibinfo{volume}{47} (\bibinfo{publisher}{American Mathematical Society
  Providence}, \bibinfo{year}{2002}).

\bibitem[{\citenamefont{Griffiths and Niu}(1996)}]{griffiths1996semiclassical}
\bibinfo{author}{\bibfnamefont{R.~B.} \bibnamefont{Griffiths}}
  \bibnamefont{and} \bibinfo{author}{\bibfnamefont{C.-S.} \bibnamefont{Niu}},
  \bibinfo{journal}{Phys. Rev. Lett.} \textbf{\bibinfo{volume}{76}},
  \bibinfo{pages}{3228} (\bibinfo{year}{1996}).

\bibitem[{\citenamefont{Parker and Plenio}(2000)}]{parker2000efficient}
\bibinfo{author}{\bibfnamefont{S.}~\bibnamefont{Parker}} \bibnamefont{and}
  \bibinfo{author}{\bibfnamefont{M.~B.} \bibnamefont{Plenio}},
  \bibinfo{journal}{Phys. Rev. Lett.} \textbf{\bibinfo{volume}{85}},
  \bibinfo{pages}{3049} (\bibinfo{year}{2000}).

\bibitem[{\citenamefont{Braginsky and Khalili}(1996)}]{braginsky1996quantum}
\bibinfo{author}{\bibfnamefont{V.~B.} \bibnamefont{Braginsky}}
  \bibnamefont{and} \bibinfo{author}{\bibfnamefont{F.~Y.}
  \bibnamefont{Khalili}}, \bibinfo{journal}{Reviews of Modern Physics}
  \textbf{\bibinfo{volume}{68}}, \bibinfo{pages}{1} (\bibinfo{year}{1996}).

\bibitem[{\citenamefont{Doherty et~al.}(2013)\citenamefont{Doherty, Manson,
  Delaney, Jelezko, Wrachtrup, and Hollenberg}}]{doherty2013nitrogen}
\bibinfo{author}{\bibfnamefont{M.~W.} \bibnamefont{Doherty}},
  \bibinfo{author}{\bibfnamefont{N.~B.} \bibnamefont{Manson}},
  \bibinfo{author}{\bibfnamefont{P.}~\bibnamefont{Delaney}},
  \bibinfo{author}{\bibfnamefont{F.}~\bibnamefont{Jelezko}},
  \bibinfo{author}{\bibfnamefont{J.}~\bibnamefont{Wrachtrup}},
  \bibnamefont{and} \bibinfo{author}{\bibfnamefont{L.~C.}
  \bibnamefont{Hollenberg}}, \bibinfo{journal}{Physics Reports}
  \textbf{\bibinfo{volume}{528}}, \bibinfo{pages}{1} (\bibinfo{year}{2013}).

\bibitem[{\citenamefont{Neumann et~al.}(2008)\citenamefont{Neumann, Mizuochi,
  Rempp, Hemmer, Watanabe, Yamasaki, Jacques, Gaebel, Jelezko, and
  Wrachtrup}}]{NMRHWYJGJW01a}
\bibinfo{author}{\bibfnamefont{P.}~\bibnamefont{Neumann}},
  \bibinfo{author}{\bibfnamefont{N.}~\bibnamefont{Mizuochi}},
  \bibinfo{author}{\bibfnamefont{F.}~\bibnamefont{Rempp}},
  \bibinfo{author}{\bibfnamefont{P.}~\bibnamefont{Hemmer}},
  \bibinfo{author}{\bibfnamefont{H.}~\bibnamefont{Watanabe}},
  \bibinfo{author}{\bibfnamefont{S.}~\bibnamefont{Yamasaki}},
  \bibinfo{author}{\bibfnamefont{V.}~\bibnamefont{Jacques}},
  \bibinfo{author}{\bibfnamefont{T.}~\bibnamefont{Gaebel}},
  \bibinfo{author}{\bibfnamefont{F.}~\bibnamefont{Jelezko}}, \bibnamefont{and}
  \bibinfo{author}{\bibfnamefont{J.}~\bibnamefont{Wrachtrup}},
  \bibinfo{journal}{Science} \textbf{\bibinfo{volume}{320}},
  \bibinfo{pages}{1326} (\bibinfo{year}{2008}).

\bibitem[{\citenamefont{Shimo-Oka et~al.}(2015)\citenamefont{Shimo-Oka, Kato,
  Yamasaki, Jelezko, Miwa, Suzuki, and Mizuochi}}]{shimo2015control}
\bibinfo{author}{\bibfnamefont{T.}~\bibnamefont{Shimo-Oka}},
  \bibinfo{author}{\bibfnamefont{H.}~\bibnamefont{Kato}},
  \bibinfo{author}{\bibfnamefont{S.}~\bibnamefont{Yamasaki}},
  \bibinfo{author}{\bibfnamefont{F.}~\bibnamefont{Jelezko}},
  \bibinfo{author}{\bibfnamefont{S.}~\bibnamefont{Miwa}},
  \bibinfo{author}{\bibfnamefont{Y.}~\bibnamefont{Suzuki}}, \bibnamefont{and}
  \bibinfo{author}{\bibfnamefont{N.}~\bibnamefont{Mizuochi}},
  \bibinfo{journal}{Appl. Phys. Lett.} \textbf{\bibinfo{volume}{106}},
  \bibinfo{pages}{153103} (\bibinfo{year}{2015}).

\bibitem[{\citenamefont{Macha et~al.}(2014)\citenamefont{Macha, Oelsner,
  Reiner, Marthaler, Andr{\'e}, Sch{\"o}n, H{\"u}bner, Meyer, llIichev, and
  Ustinov}}]{macha2014implementation}
\bibinfo{author}{\bibfnamefont{P.}~\bibnamefont{Macha}},
  \bibinfo{author}{\bibfnamefont{G.}~\bibnamefont{Oelsner}},
  \bibinfo{author}{\bibfnamefont{J.~M.} \bibnamefont{Reiner}},
  \bibinfo{author}{\bibfnamefont{M.}~\bibnamefont{Marthaler}},
  \bibinfo{author}{\bibfnamefont{S.}~\bibnamefont{Andr{\'e}}},
  \bibinfo{author}{\bibfnamefont{G.}~\bibnamefont{Sch{\"o}n}},
  \bibinfo{author}{\bibfnamefont{U.}~\bibnamefont{H{\"u}bner}},
  \bibinfo{author}{\bibfnamefont{H.~G.} \bibnamefont{Meyer}},
  \bibinfo{author}{\bibfnamefont{E.}~\bibnamefont{llIichev}}, \bibnamefont{and}
  \bibinfo{author}{\bibfnamefont{A.~V.} \bibnamefont{Ustinov}},
  \bibinfo{journal}{Nature communications} \textbf{\bibinfo{volume}{5}}
  (\bibinfo{year}{2014}).

\bibitem[{\citenamefont{Kakuyanagi et~al.}(2016)\citenamefont{Kakuyanagi,
  Matsuzaki, Deprez, Toida, Semba, Yamaguchi, Munro, and
  Saito}}]{kakuyanagi2016observation}
\bibinfo{author}{\bibfnamefont{K.}~\bibnamefont{Kakuyanagi}},
  \bibinfo{author}{\bibfnamefont{Y.}~\bibnamefont{Matsuzaki}},
  \bibinfo{author}{\bibfnamefont{C.}~\bibnamefont{Deprez}},
  \bibinfo{author}{\bibfnamefont{H.}~\bibnamefont{Toida}},
  \bibinfo{author}{\bibfnamefont{K.}~\bibnamefont{Semba}},
  \bibinfo{author}{\bibfnamefont{H.}~\bibnamefont{Yamaguchi}},
  \bibinfo{author}{\bibfnamefont{W.~J.} \bibnamefont{Munro}}, \bibnamefont{and}
  \bibinfo{author}{\bibfnamefont{S.}~\bibnamefont{Saito}},
  \bibinfo{journal}{arXiv preprint arXiv:1606.04222}  (\bibinfo{year}{2016}).

\bibitem[{\citenamefont{Heeres et~al.}(2015)\citenamefont{Heeres, Vlastakis,
  Holland, Krastanov, Albert, Frunzio, Jiang, and
  Schoelkopf}}]{heeres2015cavity}
\bibinfo{author}{\bibfnamefont{R.~W.} \bibnamefont{Heeres}},
  \bibinfo{author}{\bibfnamefont{B.}~\bibnamefont{Vlastakis}},
  \bibinfo{author}{\bibfnamefont{E.}~\bibnamefont{Holland}},
  \bibinfo{author}{\bibfnamefont{S.}~\bibnamefont{Krastanov}},
  \bibinfo{author}{\bibfnamefont{V.~V.} \bibnamefont{Albert}},
  \bibinfo{author}{\bibfnamefont{L.}~\bibnamefont{Frunzio}},
  \bibinfo{author}{\bibfnamefont{L.}~\bibnamefont{Jiang}}, \bibnamefont{and}
  \bibinfo{author}{\bibfnamefont{R.~J.} \bibnamefont{Schoelkopf}},
  \bibinfo{journal}{Phys. Rev. Lett.} \textbf{\bibinfo{volume}{115}},
  \bibinfo{pages}{137002} (\bibinfo{year}{2015}).

\bibitem[{\citenamefont{Clarke and Wilhelm}(2007)}]{ClarkeWilhelm01a}
\bibinfo{author}{\bibfnamefont{J.}~\bibnamefont{Clarke}} \bibnamefont{and}
  \bibinfo{author}{\bibfnamefont{F.~K.} \bibnamefont{Wilhelm}},
  \bibinfo{journal}{Nature} \textbf{\bibinfo{volume}{453}},
  \bibinfo{pages}{1031} (\bibinfo{year}{2007}).

\bibitem[{\citenamefont{Marcos et~al.}(2010)\citenamefont{Marcos, Wubs, Taylor,
  Aguado, Lukin, and S{\o}rensen}}]{marcos2010coupling}
\bibinfo{author}{\bibfnamefont{D.}~\bibnamefont{Marcos}},
  \bibinfo{author}{\bibfnamefont{M.}~\bibnamefont{Wubs}},
  \bibinfo{author}{\bibfnamefont{J.~M.} \bibnamefont{Taylor}},
  \bibinfo{author}{\bibfnamefont{R.}~\bibnamefont{Aguado}},
  \bibinfo{author}{\bibfnamefont{M.~D.} \bibnamefont{Lukin}}, \bibnamefont{and}
  \bibinfo{author}{\bibfnamefont{A.~S.} \bibnamefont{S{\o}rensen}},
  \bibinfo{journal}{Phys. Rev. Lett.} \textbf{\bibinfo{volume}{105}},
  \bibinfo{pages}{210501} (\bibinfo{year}{2010}).

\bibitem[{\citenamefont{Twamley and
  Barrett}(2010)}]{twamley2010superconducting}
\bibinfo{author}{\bibfnamefont{J.}~\bibnamefont{Twamley}} \bibnamefont{and}
  \bibinfo{author}{\bibfnamefont{S.~D.} \bibnamefont{Barrett}},
  \bibinfo{journal}{Phys. Rev. B} \textbf{\bibinfo{volume}{81}},
  \bibinfo{pages}{241202} (\bibinfo{year}{2010}).

\bibitem[{\citenamefont{Zhu et~al.}(2011)\citenamefont{Zhu, Saito, Kemp,
  Kakuyanagi, Karimoto, Nakano, Munro, Tokura, Everitt, Nemoto
  et~al.}}]{zhu2011coherent}
\bibinfo{author}{\bibfnamefont{X.}~\bibnamefont{Zhu}},
  \bibinfo{author}{\bibfnamefont{S.}~\bibnamefont{Saito}},
  \bibinfo{author}{\bibfnamefont{A.}~\bibnamefont{Kemp}},
  \bibinfo{author}{\bibfnamefont{K.}~\bibnamefont{Kakuyanagi}},
  \bibinfo{author}{\bibfnamefont{S.}~\bibnamefont{Karimoto}},
  \bibinfo{author}{\bibfnamefont{H.}~\bibnamefont{Nakano}},
  \bibinfo{author}{\bibfnamefont{W.~J.} \bibnamefont{Munro}},
  \bibinfo{author}{\bibfnamefont{Y.}~\bibnamefont{Tokura}},
  \bibinfo{author}{\bibfnamefont{M.~S.} \bibnamefont{Everitt}},
  \bibinfo{author}{\bibfnamefont{K.}~\bibnamefont{Nemoto}},
  \bibnamefont{et~al.}, \bibinfo{journal}{Nature}
  \textbf{\bibinfo{volume}{478}}, \bibinfo{pages}{221} (\bibinfo{year}{2011}).

\bibitem[{\citenamefont{Yan et~al.}(2015)\citenamefont{Yan, Gustavsson, Kamal,
  Birenbaum, Sears, Hover, Gudmundsen, Yoder, Orlando, Clarke
  et~al.}}]{yan2015flux}
\bibinfo{author}{\bibfnamefont{F.}~\bibnamefont{Yan}},
  \bibinfo{author}{\bibfnamefont{S.}~\bibnamefont{Gustavsson}},
  \bibinfo{author}{\bibfnamefont{A.}~\bibnamefont{Kamal}},
  \bibinfo{author}{\bibfnamefont{J.}~\bibnamefont{Birenbaum}},
  \bibinfo{author}{\bibfnamefont{A.}~\bibnamefont{Sears}},
  \bibinfo{author}{\bibfnamefont{D.}~\bibnamefont{Hover}},
  \bibinfo{author}{\bibfnamefont{T.}~\bibnamefont{Gudmundsen}},
  \bibinfo{author}{\bibfnamefont{J.}~\bibnamefont{Yoder}},
  \bibinfo{author}{\bibfnamefont{T.}~\bibnamefont{Orlando}},
  \bibinfo{author}{\bibfnamefont{J.}~\bibnamefont{Clarke}},
  \bibnamefont{et~al.}, \bibinfo{journal}{arXiv preprint arXiv:1508.06299}
  (\bibinfo{year}{2015}).

\bibitem[{\citenamefont{Tyryshkin et~al.}(2012)\citenamefont{Tyryshkin, Tojo,
  Morton, Riemann, Abrosimov, Becker, Pohl, Schenkel, Thewalt, Itoh
  et~al.}}]{tyryshkin2012electron}
\bibinfo{author}{\bibfnamefont{A.~M.} \bibnamefont{Tyryshkin}},
  \bibinfo{author}{\bibfnamefont{S.}~\bibnamefont{Tojo}},
  \bibinfo{author}{\bibfnamefont{J.~J.} \bibnamefont{Morton}},
  \bibinfo{author}{\bibfnamefont{H.}~\bibnamefont{Riemann}},
  \bibinfo{author}{\bibfnamefont{N.~V.} \bibnamefont{Abrosimov}},
  \bibinfo{author}{\bibfnamefont{P.}~\bibnamefont{Becker}},
  \bibinfo{author}{\bibfnamefont{H.-J.} \bibnamefont{Pohl}},
  \bibinfo{author}{\bibfnamefont{T.}~\bibnamefont{Schenkel}},
  \bibinfo{author}{\bibfnamefont{M.~L.} \bibnamefont{Thewalt}},
  \bibinfo{author}{\bibfnamefont{K.~M.} \bibnamefont{Itoh}},
  \bibnamefont{et~al.}, \bibinfo{journal}{Nature materials}
  \textbf{\bibinfo{volume}{11}}, \bibinfo{pages}{143} (\bibinfo{year}{2012}).

\bibitem[{\citenamefont{Reagor et~al.}(2015)\citenamefont{Reagor, Pfaff,
  Axline, Heeres, Ofek, Sliwa, Holland, Wang, Blumoff, Chou
  et~al.}}]{reagor2015quantum}
\bibinfo{author}{\bibfnamefont{M.}~\bibnamefont{Reagor}},
  \bibinfo{author}{\bibfnamefont{W.}~\bibnamefont{Pfaff}},
  \bibinfo{author}{\bibfnamefont{C.}~\bibnamefont{Axline}},
  \bibinfo{author}{\bibfnamefont{R.~W.} \bibnamefont{Heeres}},
  \bibinfo{author}{\bibfnamefont{N.}~\bibnamefont{Ofek}},
  \bibinfo{author}{\bibfnamefont{K.}~\bibnamefont{Sliwa}},
  \bibinfo{author}{\bibfnamefont{E.}~\bibnamefont{Holland}},
  \bibinfo{author}{\bibfnamefont{C.}~\bibnamefont{Wang}},
  \bibinfo{author}{\bibfnamefont{J.}~\bibnamefont{Blumoff}},
  \bibinfo{author}{\bibfnamefont{K.}~\bibnamefont{Chou}}, \bibnamefont{et~al.},
  \bibinfo{journal}{arXiv preprint arXiv:1508.05882}  (\bibinfo{year}{2015}).

\bibitem[{\citenamefont{Saeedi et~al.}(2013)\citenamefont{Saeedi, Simmons,
  Salvail, Dluhy, Riemann, Abrosimov, Becker, Pohl, Morton, and
  Thewalt}}]{saeedi2013room}
\bibinfo{author}{\bibfnamefont{K.}~\bibnamefont{Saeedi}},
  \bibinfo{author}{\bibfnamefont{S.}~\bibnamefont{Simmons}},
  \bibinfo{author}{\bibfnamefont{J.~Z.} \bibnamefont{Salvail}},
  \bibinfo{author}{\bibfnamefont{P.}~\bibnamefont{Dluhy}},
  \bibinfo{author}{\bibfnamefont{H.}~\bibnamefont{Riemann}},
  \bibinfo{author}{\bibfnamefont{N.~V.} \bibnamefont{Abrosimov}},
  \bibinfo{author}{\bibfnamefont{P.}~\bibnamefont{Becker}},
  \bibinfo{author}{\bibfnamefont{H.-J.} \bibnamefont{Pohl}},
  \bibinfo{author}{\bibfnamefont{J.~J.} \bibnamefont{Morton}},
  \bibnamefont{and} \bibinfo{author}{\bibfnamefont{M.~L.}
  \bibnamefont{Thewalt}}, \bibinfo{journal}{Science}
  \textbf{\bibinfo{volume}{342}}, \bibinfo{pages}{830} (\bibinfo{year}{2013}).

\bibitem[{\citenamefont{Higgins et~al.}(2007)\citenamefont{Higgins, Berry,
  Bartlett, Wiseman, and Pryde}}]{higgins2007entanglement}
\bibinfo{author}{\bibfnamefont{B.~L.} \bibnamefont{Higgins}},
  \bibinfo{author}{\bibfnamefont{D.~W.} \bibnamefont{Berry}},
  \bibinfo{author}{\bibfnamefont{S.~D.} \bibnamefont{Bartlett}},
  \bibinfo{author}{\bibfnamefont{H.~M.} \bibnamefont{Wiseman}},
  \bibnamefont{and} \bibinfo{author}{\bibfnamefont{G.~J.} \bibnamefont{Pryde}},
  \bibinfo{journal}{Nature} \textbf{\bibinfo{volume}{450}},
  \bibinfo{pages}{393} (\bibinfo{year}{2007}).

\bibitem[{\citenamefont{Berry et~al.}(2009)\citenamefont{Berry, Higgins,
  Bartlett, Mitchell, Pryde, and Wiseman}}]{berry2009perform}
\bibinfo{author}{\bibfnamefont{D.~W.} \bibnamefont{Berry}},
  \bibinfo{author}{\bibfnamefont{B.~L.} \bibnamefont{Higgins}},
  \bibinfo{author}{\bibfnamefont{S.~D.} \bibnamefont{Bartlett}},
  \bibinfo{author}{\bibfnamefont{M.~W.} \bibnamefont{Mitchell}},
  \bibinfo{author}{\bibfnamefont{G.~J.} \bibnamefont{Pryde}}, \bibnamefont{and}
  \bibinfo{author}{\bibfnamefont{H.~M.} \bibnamefont{Wiseman}},
  \bibinfo{journal}{Phys. Rev. A} \textbf{\bibinfo{volume}{80}},
  \bibinfo{pages}{052114} (\bibinfo{year}{2009}).

\bibitem[{\citenamefont{Yoshihara et~al.}(2016)\citenamefont{Yoshihara, Fuse,
  Ashhab, Kakuyanagi, Saito, and Semba}}]{yoshihara2016superconducting}
\bibinfo{author}{\bibfnamefont{F.}~\bibnamefont{Yoshihara}},
  \bibinfo{author}{\bibfnamefont{T.}~\bibnamefont{Fuse}},
  \bibinfo{author}{\bibfnamefont{S.}~\bibnamefont{Ashhab}},
  \bibinfo{author}{\bibfnamefont{K.}~\bibnamefont{Kakuyanagi}},
  \bibinfo{author}{\bibfnamefont{S.}~\bibnamefont{Saito}}, \bibnamefont{and}
  \bibinfo{author}{\bibfnamefont{K.}~\bibnamefont{Semba}},
  \bibinfo{journal}{arXiv preprint arXiv:1602.00415}  (\bibinfo{year}{2016}).

\end{thebibliography}

\end{document}